\documentclass[aps,prl,twocolumn,groupedaddress,showpacs]{revtex4}
\usepackage{epsfig,amssymb}
\begin{document}
{\scriptsize Phys. Rev. Lett. {\bf 88}, 183602 (2002).}
\title{Generating Entangled Two-Photon States with Coincident
Frequencies}
\author{Vittorio Giovannetti, Lorenzo Maccone, Jeffrey  H. Shapiro,
and Franco N. C. Wong}
\affiliation{Massachusetts Institute of Technology, Research
Laboratory of Electronics, Cambridge MA 02139, USA}

\date{\today}
\begin{abstract}
It is shown that parametric downconversion, with a short-duration pump
pulse and a long nonlinear crystal that is appropriately phase
matched, can produce a frequency-entangled biphoton state whose
individual photons are coincident in frequency.  Quantum interference
experiments which distinguish this state from the familiar
time-coincident biphoton state are described.
\end{abstract}

\pacs{03.65.Ud, 42.50.Dv, 03.67.-a, 42.65.-k}
\maketitle

Spontaneous parametric downconversion (SPDC) has been the entanglement
source of choice for experimental demonstrations of quantum
teleportation, entanglement-based quantum cryptography, and
Bell-inequality violations, {\it etc}.  However, the biphoton state
generated via SPDC under the customary phase-matching conditions is
maximally entangled only when a continuous-wave (cw) pump is used
{\cite{theory,wal}}.  In pulsed-pump experiments, the fringe
visibility in biphoton interference measurements decreases as the
duration of the pump pulse is reduced {\cite{keller}}. The timing and
pump-intensity advantages of pulsed experiments has thus spurred work
to retain or restore maximal entanglement in the pulsed regime
{\cite{altri}}.

In this paper a new method for obtaining maximal entanglement from
pulsed SPDC is reported.  Our approach uses a long nonlinear crystal
and extended phase-matching conditions tailored specifically to pulsed
operation.  As such, it does not require any filtering or
post-selection, and it reaps the high-conversion-efficiency advantage
that long crystals afford.  Furthermore, the biphoton states that it
produces are comprised of photons that are coincident in frequency, in
contrast to the usual cw phase-matching case whose biphotons exhibit
coincidence in time. Coincident-in-frequency entanglement is important
because the $N$-photon version of such a state has been shown to
improve the accuracy of time-of-flight position sensing or clock
synchronization by a factor of $\sqrt{N}$ {\cite{paper}}.

Consider SPDC with a cw pump and conventional phase matching that is
operated at frequency degeneracy, {\it i.e.}, phase matched such that
the center frequencies of its signal and idler equal half the pump
frequency.  This system produces a biphoton of the form,
\begin{eqnarray} |{\rm TB}\rangle\equiv\int
\frac{d\omega}{2\pi}\;\phi(\omega)|\omega_p/2-\omega\rangle_s
|\omega_p/2+\omega\rangle_i
\;\label{tbst}.
\end{eqnarray}
Here: $|\omega_s\rangle_s$ and $|\omega_i\rangle_i$ are single-photon
signal and idler states in which the photons are present at
frequencies $\omega_s$ and $\omega_i$, respectively; $\omega_p$ is the
pump frequency; and $\phi(\omega)$ is the spectral function of the
state, so that $|\phi(\omega_p/2-\omega)|^2$ is the signal's
fluorescence spectrum.  The notation $|{\rm TB}\rangle$ indicates that
Eq.~(\ref{tbst}) is the usual twin-beam state of SPDC.  The frequency
entanglement of this state dictates that a signal photon at frequency
$\omega_p/2-\omega$ is accompanied by an idler photon at frequency
$\omega_p/2 +\omega$.  The sum of the signal and idler frequencies is
therefore fixed at the pump frequency.  By Fourier duality, this
implies that the signal and idler photons occur in time
coincidence---to within a reciprocal fluorescence bandwidth---as has
been shown in the famous ``Mandel dip'' experiment {\cite{mandel_dip}}.

On the other hand, the SPDC biphoton
that will be studied in this paper is,
\begin{eqnarray}
|{\rm DB}\rangle=\int \frac{d\omega}{2\pi}\;
\phi(\omega)|\omega_p/2+\omega\rangle_s |\omega_p/2+\omega\rangle_i
\label{ntbst}.
\end{eqnarray}
In this state, a signal photon at $\omega_p/2+\omega$ is accompanied
by an idler photon at the same frequency.  This
coincident-in-frequency behavior leads, via Fourier duality, to
symmetrically located occurrences in time.  In particular, a signal
photon appearing at $T_0 + t$ is accompanied by an idler photon at
$T_0 - t$, where $T_0$ is the mean time-of-arrival of the biphoton
pulse.  Because this biphoton possesses a narrow distribution in
signal-minus-idler difference frequency, we have dubbed it the
difference-beam ($|{\rm DB}\rangle)$ state.  Note that its mean
time-of-arrival, $T_0$, plays the role of the fluorescence center
frequency, $\omega_p/2$, in comparing the DB and TB states.  Thus,
whereas the photons in $|{\rm TB}\rangle$ may be discriminated by
frequency measurements, those in $|{\rm DB}\rangle$ may be
distinguished via time-of-arrival measurements.

The rest of the paper is organized as follows.  First, we derive the
output state of SPDC. The phase-matching conditions that are needed to
create the DB state are then obtained and explained.  Next, we present
quantum interference experiments that can distinguish between the TB
and DB states.  Finally, we give a feasibility study for $|{\rm
DB}\rangle$ generation using periodically-poled potassium titanyl
phosphate (PPKTP).

A textbook treatment of the SPDC process (see for example
{\cite{mandel}}) allows us to deduce the state at the output of a
compensated SPDC crystal.  We will give a brief derivation here
assuming colinear plane-wave operation.  In the interaction picture
under the rotating-wave approximation, the Hamiltonian that gives rise
to the creation of the two downconverted photons starting from a
single pump photon is given by,
\begin{eqnarray}
\lefteqn{H_I(t)=}\label{hamilt}
\\\nonumber
& & S\int_{-L/2}^{L/2}\!dz\,\chi^{(2)}E_p^{(+)}(z,t) E_s^{(-)}(z,t)
E_i^{(-)}(z,t) + {\rm h.c.},
\end{eqnarray}
where $\chi^{(2)}$ is the nonlinear coefficient and $L$ is the length
of the downconversion crystal, $S$ is the pump-beam area, and
$E^{(+)}$ and $E^{(-)}\equiv (E^{(+)})^\dagger$ are positive-frequency
and negative-frequency electric field operators with the subscripts
$\{p,s,i\}$ denoting pump, signal, and idler, respectively.  These
electric field operators obey,
\begin{eqnarray} 
{E_j^{(+)}(z,t) = i\!\int\!
\frac{ d\omega}{2\pi}\,\sqrt{\frac{\pi\
\hbar\omega}{c\epsilon_0n_j^2(\omega)S}} 
\,a_j(\omega)\, e^{i[k_j(\omega)z-\omega t]},}
\end{eqnarray}
for $j = p,s,i$, where $a_j(\omega)$ is the annihilation operator for
frequency-$\omega$ photons, $n_j(\omega)$ is the refractive index for
the $j$th beam (pump, signal, or idler), and $k_j(\omega)\equiv \omega
n_j(\omega)/c$ is the associated wave number.

The Hamiltonian (\ref{hamilt}) yields the state at the output of the
crystal (for vacuum-input signal and idler) via,
\begin{equation}
|\Psi\rangle \simeq |0\rangle -\frac i{\hbar}\int_{t_0}^t\!
dt'\,H_I(t')|0\rangle,
\label{evol}
\end{equation}
for small values of the coupling constant $\chi^{(2)}$.  As done in
{\cite{wal}}, we shall assume that $\chi^{(2)}$ is independent of
frequency over the pump bandwidth, even though this assumption may not
be satisfied in some ultrafast applications.  For a strong coherent
pump pulse and in the absence of pump depletion, we may replace the
pump field operator in Eq.~(\ref{hamilt}) with,
\begin{equation}
E_p^{(+)}(z,t)\simeq \int\! \frac{d\omega} {2\pi}\, {\cal
E}_p(\omega)\, e^{i[k_p(\omega)z-\omega t]},
\label{campocl}
\end{equation}
where ${\cal E}_p(\omega)$ is a classical complex amplitude.  Because
we are interested in the fields far from the crystal, we may expand
the integration limits in Eq. (\ref{evol}) to run from $-\infty$ to
$+\infty$. Thus, the $t'$ integration produces an impulse,
$\delta(\omega_p-\omega_s-\omega_i)$, that expresses energy
conservation at the photon level.

The biphoton state that we are seeking is the non-vacuum part of
Eq.~(\ref{evol}).  Under the preceding assumptions, this is given by,
\begin{eqnarray}
\lefteqn{|\Psi\rangle =} \label{stato}
\\ \nonumber
& & i\frac{\chi^{(2)}\pi}{c\epsilon_0}
\int\! \frac{d\omega_s}{2\pi}
\int\! \frac{d\omega_i}{2\pi}\;
\alpha(\omega_s,\omega_i)
\Phi_L(\omega_s,\omega_i)\;|\omega_s\rangle_s|\omega_i\rangle_i,
\end{eqnarray}
where $|\omega\rangle\equiv a^\dag(\omega)|0\rangle$ is a single-photon state,
\begin{equation}
\alpha(\omega_s,\omega_i)\equiv\frac{\sqrt{\omega_s\omega_i}}
{n_s(\omega_s)n_i(\omega_i)}\,
{\cal E}_p(\omega_s+\omega_i),
\label{alpha}
\end{equation}
is determined by the pump spectrum, and
\begin{equation}
\Phi_L(\omega_s,\omega_i) \equiv
\frac{\sin\left(\Delta k(\omega_s,\omega_i)L/2\right)}
{\Delta k(\omega_s,\omega_i)/2},
\label{defphi}
\end{equation}
is the phase-matching function, with $\Delta k(\omega_s,\omega_i)\equiv
k_p(\omega_s+\omega_i)-k_s(\omega_s)-k_i(\omega_i)$.

To obtain maximal entanglement from the biphoton state (\ref{stato})
we need to collapse the double integral over frequency into a single
integral.  For the customary twin-beam state $|{\rm TB}\rangle$, this
is accomplished by using a cw pump of frequency $\omega_p$ to force
$\alpha(\omega_s,\omega_i)\propto\delta(\omega_s+\omega_i-\omega_p)$
in Eq. (\ref{stato}).  We then obtain a TB state (1) with spectral
function
$\phi(\omega)\propto\Phi_L(\omega_p/2-\omega,\omega_p/2+\omega)$.
Because this makes the common signal/idler fluorescence bandwidth,
$\Omega_f$, inversely proportional
to $L$, it follows that short
crystals are better suited to generating broadband TB states. For DB
states, however, we will see that long
crystals do not prevent broadband entanglement generation.

Continuous-wave operation is not the only way to obtain a
maximally-entangled state from (\ref{stato}). We can also eliminate
one of the frequency integrals by forcing $\Phi_L$ to approach a delta
function. The property $\lim_{L\to\infty}[\sin (xL)/x]=\pi\delta(x)$
allows us to write $\Phi_L(\omega_s,\omega_i)=2\pi\delta\left (\Delta
k(\omega_s,\omega_i)\right)$ for an infinitely long crystal. (In
practice the nonlinear crystal will always have a finite length $L$,
but we will see that a high degree of entanglement can be obtained
using practical values of $L$.)  To force
$\Phi_L(\omega_s,\omega_i)\propto\delta(\omega_s-\omega_i)$ in the
long-crystal limit, we need to ensure that $\Delta
k(\omega_s,\omega_i)=0$ if and only if $\omega_s=\omega_i$, for
$\omega_s+\omega_i$ ranging over the full pump bandwidth, $\Omega_p$.
Equation~(\ref{stato}) then reduces to the DB state of
Eq.~(\ref{ntbst}), with spectral function
$\phi(\omega)=\alpha(\omega_p/2+\omega,\omega_p/2+\omega)$, where
$\omega_p$ is the pump beam's center frequency. Note that
$\phi(\omega)$ depends only on the pump spectrum and the refractive
indexes of the nonlinear crystal, as can be seen from
Eq.~(\ref{alpha}), and that its bandwidth is $\Omega_p/2$. Moreover,
the symmetry of the phase-matching function $\Phi_L$ forces the signal
and idler fluorescence spectra to be identical, something that is not
generally true in ultrafast type-II downconversion {\cite{wal}}.

Is it possible to satisfy the condition $\Delta
k(\omega_s,\omega_i)=0$ only for $\omega_s=\omega_i$ over the full
pump bandwidth? What does this condition correspond to physically?  By
using the first-order Taylor expansions of $k_s$ and $k_i$ around
$\omega_p/2$ and of $k_p$ around $\omega_p$, we find that,
\begin{eqnarray}
n_p(\omega_p)
&=&\frac{n_s(\omega_p/2)+n_i(\omega_p/2)}2,
\label{ordine0}\\
k'_p(\omega_p) &=&
\frac{k'_s(\omega_p/2)+k'_i(\omega_p/2)}2,\label{ordine1}
\end{eqnarray}
ensure that $\Delta k(\omega_p/2+\omega,\omega_p/2+\omega) = 0$ for
$|\omega|\le \Omega_p/2$.  In physical terms, the extended
phase-matching condition given by (\ref{ordine0}) and (\ref{ordine1})
assert that the index of refraction and the inverse group velocity
seen by the pump must equal the averages of those seen by the signal
and idler. Equation~(\ref{ordine0}) is the customary phase-matching
condition required for the generation of $|{\rm TB}\rangle$ at
frequency degeneracy.  Equation~(\ref{ordine1}) is equivalent to the
``group velocity matching'' condition introduced in {\cite{keller}}.
It turns out, however, that Eqs.~(\ref{ordine0}) and (\ref{ordine1})
are not sufficient for DB state generation.  Because $\Delta
k(\omega_s,\omega_i)$ must vanish only for $\omega_s=\omega_i$, we
must also require that $k'_s(\omega_p/2)\neq k'_i(\omega_p/2)$.  This
requirement excludes type-I crystals, for which $k_s(\omega) =
k_i(\omega)$.  Thus, in all that follows we will presume type-II
operation. We will discuss later the validity of truncating the Taylor
series at the $n=1$ terms.

\begin{figure}[hbt]
\begin{center}\epsfxsize=.75
\hsize\leavevmode\epsffile{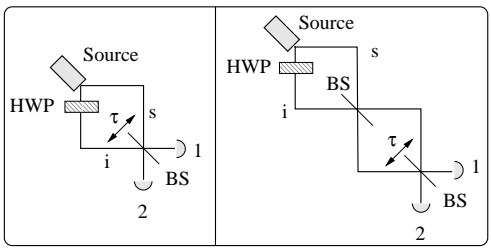}
\end{center}
\caption{Quantum interference experiments to distinguish $|{\rm
DB}\rangle$ from $|{\rm TB}\rangle$.  The left panel shows the HOM
interferometer and the right shows the MZ interferometer.  In both
cases, coincidence counts are measured as the relative delay, $\tau$,
between the interferometer's arms is varied by moving the beam
splitter that is nearest to the detectors. The half-wave plate (HWP)
rotates the idler polarization to match that of the signal because
type-II downconversion is assumed.}
\label{f:hom}\end{figure}

The states $|{\rm DB}\rangle$ and $|{\rm TB}\rangle$ are duals in the
following sense.  The former is a biphoton whose constituent photons
are coincident in frequency, and the latter is a biphoton whose
constituent photons are time-coincident.  We now show that coincidence
counting using Hong-Ou-Mandel (HOM) and Mach-Zehnder (MZ)
interferometers, as sketched in Fig.~{\ref{f:hom}}, can provide
experimental quantum-interference signatures that distinguish between
the TB and DB biphoton states.

To understand the outcome of the two experiments shown in
Fig.~{\ref{f:hom}}, it is useful to start from the general state
$|\Psi\rangle$ of the form (\ref{stato}). The coincidence rate for
detection intervals that are long compared to the reciprocal
fluorescence bandwidth is given by (see for example
{\cite{theory,wal}}),
\begin{equation}
P\propto \int \!\frac{d\omega_1}{2\pi}
\int \!\frac{d\omega_2}{2\pi}\,|\langle
0|a_1(\omega_1)
a_2(\omega_2)|\Psi\rangle|^2,
\label{rate}
\end{equation}
where $a_1$ and $a_2$ are the photon annihilation operators at the two
detectors and $|\Psi\rangle$ is the biphoton state of the source.
Assume that the product
$\alpha(\omega_1,\omega_2)\Phi_L(\omega_1,\omega_2)$ is symmetric in
$\omega_1,\omega_2$, as is the case for both $|{\rm TB}\rangle$ and
$|{\rm DB}\rangle$. By applying the beam-splitter transformations on
the operators $a_s$ and $a_i$, we find that,
\begin{eqnarray}
P_{\pm}(\tau)&\propto&\int \! \frac{d\omega_1}{2\pi}\int \!
\frac{d\omega_2}{2\pi}\,|\alpha(\omega_1,\omega_2)\;
\Phi_L(\omega_1,\omega_2)|^2\nonumber\\&\times &
\Big(1\pm
\cos[(\omega_1\pm\omega_2)\tau]\Big).
\label{biancaneve}
\end{eqnarray}
In (\ref{biancaneve}) the minus signs apply to the HOM
interferometer, and the plus signs apply to the MZ interferometer.

For the TB state, we set
$|\alpha(\omega_1,\omega_2)|^2\propto\delta(\omega_p-\omega_1-\omega_2)$
and
$|\Phi_L(\omega_1,\omega_2)|^2\propto|\phi[(\omega_2-\omega_1)/2]|^2$.
Approximating the fluorescence spectrum by
$|\phi(\omega)|^2=\sin^2(2\pi\omega/\Omega_f)/[2\pi\omega
/(\Omega_fL)]^2$, where $\Omega_f\equiv
4\pi/(L|k'_s(\omega_p/2)-k'_i(\omega_p/2)|)$, Eq.~(\ref{biancaneve})
then yields the familiar triangular-shaped HOM coincidence dip of
width $4\pi/\Omega_f$ centered at $\tau=0$ {\cite{theory,wal}}. The
$|{\rm TB}\rangle$ coincidences require that one photon exits from
each output port of the beam splitter. At zero relative delay, the two
quantum trajectories that give rise to such coincidences destructively
interfere, leading to a coincidence null {\cite{pittman}}. The width
of this dip is $\sim$1$/\Omega_f$, because signal and idler wave
packets separated by many reciprocal fluorescence bandwidths are
distinguishable and hence do not interfere.  When the TB-state
coincidence rate is evaluated for the MZ interferometer, we obtain
$P_+(\tau)\propto 1+\cos(\omega_p\tau)$, {\it i.e.}, sinusoidal
fringes at the pump frequency.  These fringes have infinite extent
because a perfect cw pump has infinite coherence time.

Now suppose that the input state in Fig.~\ref{f:hom} is $|{\rm
DB}\rangle$, {\it i.e.}, let
$|\Phi_L(\omega_1,\omega_2)|^2\propto\delta(\omega_1-\omega_2)$ and
$|\alpha(\omega_1,\omega_2)|^2=|\phi[(\omega_1+\omega_2-\omega_p)/2]|^2$
in Eq.~(\ref{biancaneve}). In this case the frequency coincidence
between the signal and idler photons eliminates any delay dependence
in the HOM configuration, reducing Eq.~(\ref{biancaneve}) to
$P_-(\tau) = 0$. In fact, the wavefunctions for the two photons extend
to all times and cannot be separated: the quantum trajectories that
give rise to coincidences destructively interfere for any delay
$\tau$. For the Mach-Zehnder arrangement, the DB state gives
$P_+(\tau)\propto 1+\exp(-\Omega_p^2\tau^2/4)\cos(\omega_p\tau)$,
under the assumption of Gaussian pump spectrum
$|\phi(\omega)|^2\propto\exp[-4\omega^2/\Omega_p^2]$.  Notice that in
this case $P_+$ again exhibits pump-frequency interference fringes,
but now the interference pattern has width $4/\Omega_p$, {\it i.e.},
roughly equal to the duration of a transform-limited pump
pulse. Similar interference patterns have been previously analyzed in
{\cite{jeff1}}.

Both the HOM and the MZ interferometers distinguish between the states
$|{\rm DB}\rangle$ and $|{\rm TB}\rangle$.  However, because DB state
generation requires infinite crystal length---whereas TB state
generation uses a finite-length crystal---it behooves us to study what
happens in the finite-$L$ regime when we use a pulsed pump in
conjunction with our extended phase-matching conditions.  The biphoton
state, $|{\rm DB}_L\rangle$, that this system generates is entangled
in frequency, but not maximally so, {\it i.e.}, measuring the
frequency of the signal photon does not exactly determine the
frequency of the idler photon.  When $|{\rm DB}_L\rangle$ is measured
with an HOM interferometer, the resulting coincidence null is no
longer of unlimited extent.  Indeed, the width of the coincidence dip
for $|{\rm DB}_L\rangle$ is identical to that for $|{\rm TB}\rangle$.
Thus the HOM interferometer cannot distinguish between these two
states.  The MZ interferometer, however, does distinguish between
$|{\rm DB}_L\rangle$ and $|{\rm TB}\rangle$, as the width of the
former's fringe pattern is set by the pump bandwidth and hence
independent of crystal length.

More insight into the complementary behavior of $|{\rm TB}\rangle$ and
$|{\rm DB}\rangle$ can be gained by examining their time domain
structures.  Both of these biphoton states arise from the coherent
superposition
of spatially-localized, instantaneous signal/idler pair creations occurring
throughout the length of the
nonlinear  crystal.  HOM and MZ interferometers use integrating
photodetectors, but reveal
temporal aspects of $|{\rm TB}\rangle$ and $|{\rm DB}\rangle$ via quantum
interference.
Suppose, however, that we use an ultrafast photodetector to measure the
arrival time
of the signal photon.  This measurement specifies a definite location,
along the crystal, at
which  the detected signal photon was created, and implies rather different
temporal statistics for its associated idler photon depending on whether
the biphoton was
$|{\rm TB}\rangle$ or $|{\rm DB}\rangle$.  Because the TB state is produced
by a cw pump,
its component photons may be created at any time.  However, once its signal
photon has been detected at
time $T_s$, the accompanying idler photon must be at $T_i$, where
$|T_s-T_i|\le\ 4\pi/\Omega_f$ for
our type-II system.  The individual photons in the DB
state  also may be created at any time, even though this biphoton is
generated by a pulsed pump.  Here, the
timing uncertainty is really uncertainty in the location, within the
infinitely-long crystal,
at which the photon pair is generated.  Once again, detection of a signal
photon at time $T_s$ provides
location information which strongly constrains the arrival time for the
idler photon.  In particular, the
extended phase-matching conditions that produce the DB state under pulsed
pumping force
$(T_s+T_i)/2$ to have a mean value at a fixed offset---set by
dispersion---from the peak of the classical
Gaussian pump pulse.

It turns out to be difficult to find a crystal satisfying the two
conditions (\ref{ordine0}) and (\ref{ordine1}).  Thus, we will enforce
(\ref{ordine0}) via quasi-phase-matching in a periodically-poled
$\chi^{(2)}$ material {\cite{qpm}}, {\it i.e.}, one for which the
addition of an artificial grating results in a spatially-varying
nonlinear coefficient, $\chi^{(2)}(z)=\chi^{(2)}\exp(i2\pi
z/\Lambda)$, along the propagation axis.  By choosing the grating
period $\Lambda$ to cancel the zeroth-order term in the $\Delta
k(\omega_s,\omega_i)$ expansion, we can replace Eq.~(\ref{ordine0})
with the new condition $n_p(\omega_p) =
[{n_s(\omega_p/2)+n_i(\omega_p/2)}]/2-{2\pi
c}/(\Lambda{\omega_p})$. This, together with Eq.~(\ref{ordine1}) is
satisfied by PPKTP at a pump wavelength of 790\,nm with a grating
period of 47.7\,$\mu$m when propagation is along the crystal's $X$
axis, the pump and idler are $Y$-polarized, and the signal is
$Z$-polarized.  It still remains for us to examine the validity
conditions for the $L\rightarrow\infty$ approximation to the
phase-matching function $\Phi_L(\omega_s,\omega_i)$.  These can be
shown to be $2\pi/\gamma\Omega_p \ll L \ll 8\pi/\mu\Omega_p^2,$ where
$\gamma\equiv |k'_p(\omega_p) - k'_s(\omega_p/2)|$, and $\mu$ is the
maximum-magnitude eigenvalue of the Hessian matrix associated with the
2-D Taylor series expansion of $\Delta k(\omega_s,\omega_i)$.
Physically, the lower limit on crystal length is set by our need to be
in the long-$L$ regime, and the upper limit is set by the second-order
terms in the Taylor expansion.  For our PPKTP example, we have that
$\gamma \approx 1.4\times 10^{-4}\,$ps/$\mu$m and $\mu \approx
3.6\times10^{-7}\,$ps$^2$/$\mu$m.  With a 170\,fs ($\Omega_p/2\pi =
3\,$THz) transform-limited pump pulse the preceding crystal-length
restrictions reduce to $0.23\,{\rm cm}\ll L \ll 19.7\,{\rm cm}$, so
that a 2-cm-long crystal will suffice. Finally, we note that
polarization-entangled DB states can be created by paralleling the
procedure in {\cite{jeff}} for the creation of polarization-entangled
TB states.

This work was funded by the ARO under a MURI program (Grant DAAD
19-00-1-0177) and by the NRO.

\end{document}